# Fine structure of perturbed Laguerre-Gaussian beams: Hermite-Gaussian mode spectra and topological charge

A. Volyar[1]*, E. Abramochkin[2], Yu. Egorov1, M. Bretsko[1], Ya. Akimova[1]

[1] V.I. Vernadsky Crimean Federal University, Vernadsky Prospect, 4, Simferopol, 295007, Russia.
[2] Samara Branch of P.N. Lebedev Physical Institute of Russian Academy of Sciences, Samara, Russia
*Corresponding author: volyar@singular-optics.org

Abstract

We found that small perturbations of the optical vortex core in the Laguerre-Gaussian (LG) beams generate a fine structure of the Hermite-Gauss (HG) mode spectrum. Such perturbations can be easily simulated by weak variations of amplitudes and phases of the HG modes in the expansion of the LG beam field. We also theoretically substantiated and experimentally implemented a method for measuring the topological charge of LG beams with an arbitrary number of ring dislocations. Theoretical discussion and experimental studies were accompanied by simple examples of estimating the orbital angular momentum and the topological charge of perturbed LG beams.

## 1. Introduction

Despite a wide application of Laguerre-Gaussian beams (LG) not only in optical devices, but also in areas of human activity far from laser optics [1-4], they turn out to be extremely unstable wave structures to numerous external perturbations. The unique property of LG beams bearing optical vortices with a large orbital angular momentum (OAM) is smoothed by a weak perturbation that turns the axial optical vortex with a large topological charge (TC) into many single-charged vortices covering the entire beam cross-section alongside with reducing both the TC and OAM [5]. But the LG beams structure is also shaped by degenerate ring dislocations so that their perturbation together with high-order vortices entails occurring a necklace of links and knots of optical vortices at the vicinity of LG beam waist [6]. In fact, in a real optical experiment, we are dealing not with single LG beams, but with complex wave combinations of vortex beams, and revealing their fine structure requires appropriate approaches and techniques. The fine structure of the perturbed LG beam is formed by wave modes, the type of which is specified by the geometric shape and optical characteristics of the disturbing object. For instance, the perturbation of the LG beam by a circular hard-edged aperture generates a set of secondary LG modes with the same TC l=const but different radial numbers n [7] while the sector aperture vice-versa gives rise to the spectrum of LG modes with the same radial numbers n=const but different TC l [8], and the modes can have different both amplitudes and initial phases [9]. Thus, the fine structure detection techniques should, on the one hand, not destroy the beam structure measuring the amplitude and phase mode spectra, but, on the other hand, sort the modes by their quantum numbers taking into account their functional basis.

The spatial sorting of LG beams by a topological charge l by means of holographic gratings was proposed back in early 2000th [10,11], while the authors of Ref. [12] address an elegant a multi-plane light conversion system that provides a unitary conversion of an array of Gaussian beams into HG modes with further astigmatic transform and sorting into the LG beam array both by the TC l and a radial number n. A detailed analysis of HG mode sorting via a fractional Fourier transform is presented in Ref. [13]. However, these techniques rely on breaking down the complex beam structure that is accompanied by losses of information about the initial phase and, therefore, the inability to restore the initial singular beam while the intensity moments technique proposed in Ref. [14] enables one to avoid the beam destruction and to measure the amplitudes and phases spectra of LG beams.

An essential part in distorting the LG beam structure is played by off-axis perturbations such as, for example, a shift of a circular hard-edged aperture, cutting the beam with an optical knife or astigmatism of optical elements (see e.g.[5,15] and references therein). A fine structure of such perturbed beams is essentially shaped by HG modes. This means that the perturbation violates the internal coupling of the HG mode in the composition of a single LG beam [16,17], i.e. small variations of the amplitudes and phases of the HG modes nucleate crucial traits in a high-order LG beam that can distort the beam structure beyond recognition. It is this problem that we are discussing in the presented article. Thus, the aim of our article is the theoretical justification and experimental implementation of the intensity moments technique for measuring the amplitude and initial phases spectra of the perturbed LG beam, and also the calculation of the TC and OAM via spectral analysis.

## II Vortex topological charge

The authors of Ref. [18] proposed and implemented a relatively simple and reliable technique for measuring topological charge (TC) $l = n - m$, $n, m = 0, 1, 2, \ldots$ by means of astigmatic transformation of LG beams with zero radial number $n = 0$ using a single cylindrical lens. Based on the solution of the diffraction problem, it was shown that a field intensity distribution corresponding to the HG mode with a complex argument is formed in the plane of the double focus f of the cylindrical lens. Along the axes of astigmatism $\varphi = 135°$ (for positive TC $l > 0$) and $\varphi = 45°$ (for negative TC $l < 0$) the number of zeros of the HG beam is equal to the topological charge of the LC beam It is worth noting that the argument of the Hermite polynomial of complex beam amplitude becomes a real at the throughout plane z=2f under the condition $2f = z_0$, where $z_0 = k w_0^2 / 2$ stands for the Rayleigh length, $w_0$ is the waist radius of the Gaussian beam at the plane of the cylindrical lens z=0, $k$ is the wavenumber. In this section, we extend this approach to the arbitrary case of a radial number $n \geq 0$. First of all, we note that in the astigmatic system, the Laguerre-Gaussian modes or the Hermite-Gauss modes are restored only at special cross sections of the laser beam. In the rest field area, it is customary to talk about orthogonal states in the form of Hermite – Laguerre – Gaussian (HLG) modes [19]. The evolution of the HLG beam between the initial state of a LG beam at the cylindrical lens z=0 and its conversion to the HG beam at the double focus plane z=2f is shown by the example of the intensity distributions in Fig. 1 (see also Section III) obtained at the experiment.

To describe the measuring process of the TC, we first write the complex amplitudes of the LG and HG beams

$$LG_{n,\pm \ell}(\mathbf{r}) = \exp(-r^2) r^\ell e^{\pm i\ell\varphi} L_n^\ell(2r^2),$$
$$HG_{n,m}(\mathbf{r}) = \exp(-r^2) H_n(\sqrt{2}x) H_m(\sqrt{2}y),$$
(1)

where $n \geq 0$, $m \geq 0$, $l \geq 0$, $\mathbf{r}=(x,y)=(r\cos\varphi, r\sin\varphi)$, $\rho, \varphi$ are the polar coordinates, $r = \rho/w_0$ stands for a dimensionless radial variable, $L_n^l(2r^2)$ and $H_n(\sqrt{2}x)$ are Laguerre and Hermite polynomials, respectively. Hereinafter, we use dimensionless coordinates for simplicity. The transition to dimensional coordinates is carried out by replacing $(X,Y,Z) = (w_0 x, w_0 y, z_0 z)$.

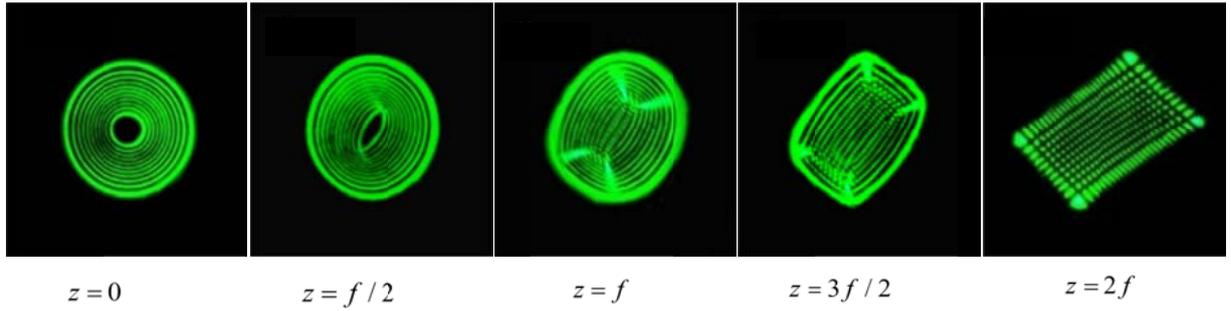

$z=0$    $z=f/2$    $z=f$    $z=3f/2$    $z=2f$

*Fig. 1. Astigmatic transform of a LG beam with a topological charge $l=-15$ and a radial number $n=10$ of the cylindrical lens along a double focal length (experiment), f=12cm*

It is known [19] (see also Ref. [20]) that HG and LG beams are partial representatives of the parametric family of $HLG(\mathbf{r}|\alpha)$ beams for the parameter $\alpha$ takes values 0, $\pi/4$ and $\pi/2$:

$$HLG_{n,m}(\mathbf{r}|0) = (-i)^m HG_{n,m}(\mathbf{r}),$$
$$HLG_{n,m}(\mathbf{r}|\pi/2) = i^n HG_{m,n}(\mathbf{r}),$$
$$HLG_{n+m,n}(\mathbf{r}|\pi/4) = (-1)^n 2^{n+m} n!\, LG_{n,m}(\mathbf{r}),$$
$$HLG_{n,n+m}(\mathbf{r}|\pi/4) = (-1)^n 2^{n+m} n!\, LG_{n,-m}(\mathbf{r}).$$
(2)

One can combine the last two equalities into one:

$$HLG_{n,m}(\mathbf{r}|\pi/4) = (-1)^{\min} 2^{\max} (\min)!\, LG_{\min, n-m}(\mathbf{r}),$$
(3)

where $\min = \min(n,m)$, $\max = \max(n,m)$. Thus, the sign and value of the TC are determined by the difference in the indices of the HLG mode: $l = n - m$.

We now consider the LG beam (3) in the plane $z = 0$ of a cylindrical lens [19]. Right away after the astigmatic element, it takes the form

$$HLG_{n,m}^{(ast)}(\mathbf{r}, z = 0 \mid \pi/4) = $$
$$= \exp\left(i[c_x x^2 + c_y y^2]\right) HLG_{n,m}(\mathbf{r} \mid \pi/4). \quad (4)$$

The parameters $c_x$ and $c_y$ characterize the astigmatic element, and in our case, when the axis of astigmatism is oriented along the x axis, they are written as

$$c_x = -z_0/f = -2, \qquad c_y = 0. \quad (5)$$

The HLG beam propagation (4) along the z axis is obtained using the Fresnel transform:
$$\Psi_{n,m}(\mathbf{r}, z) = \mathbf{FR}_z \left[ HLG_{n,m}^{(ast)}(\boldsymbol{\rho}, z = 0 \mid \pi/4) \right].$$

We use the expression (18) from Ref.[19] and, omitting the intermediate calculations, immediately write down the answer:

$$\Psi_{n,m}(\mathbf{r}, z) = \frac{\exp\left[i\psi(\mathbf{r}) + i\psi_0\right] HLG_{n,m}(\mathbf{r}_1 \mid \pi/4 - \omega)}{\sqrt{|\sigma_x \sigma_y|}}, \quad (6)$$

where

$$\sigma_x = 1 + (c_x + i)z, \qquad \sigma_y = 1 + (c_y + i)z,$$

$$\mathbf{r}_1 = \frac{1}{\sqrt{2}} \left( \frac{x}{|\sigma_x|} + \frac{y}{|\sigma_y|}, \frac{y}{|\sigma_y|} - \frac{x}{|\sigma_x|} \right),$$

$$\psi(\mathbf{r}) = \frac{c_x + (1 + c_x^2)z}{|\sigma_x|^2} x^2 + \frac{c_y + (1 + c_y^2)z}{|\sigma_y|^2} y^2,$$

$$\psi_0 = (n - m)\frac{\pi}{4} - (n + m + 1)\frac{\arg \sigma_x + \arg \sigma_y}{2},$$

$$\omega = \left(\arg \sigma_y - \arg \sigma_x\right)/2. \quad (7)$$

In our case, the observation plane coincides with the double focus plane of the cylindrical lens, $Z = 2f$, therefore $z = 2f/z_0 = 1$ and

$$\sigma_x = \sqrt{2} e^{i3\pi/4}, \qquad \sigma_y = \sqrt{2} e^{i\pi/4},$$
$$\mathbf{r}_1 = (x + y, y - x)/2, \qquad \psi(\mathbf{r}) = (3x^2 + y^2)/2,$$
$$\psi_0 = -\pi(n + 3m + 2)/4, \qquad \omega = -\pi/4. \quad (8)$$

Substituting the parameters found in Eq. (6) and returning from the HLG modes to LG and HG modes, we find that the LG beam in the double focus plane of a cylindrical lens turns into an HG beam

$$LG_{n,\pm l}(x, y) \to \frac{i^{n-1} e^{-i\pi(1 \mp 2l/4)}}{2^{n+l+1/2} n!} \times$$
$$\times \exp\left(i \cdot \frac{3x^2 + y^2}{2}\right) HG_{n+\ell,n}\left(\frac{y \mp x, y \pm x}{2}\right). \quad (9)$$

In the interval between the initial plane and the double focus plane of the cylindrical lens, HLG states are formed. The transformation process of HLG modes is illustrated by the experimental intensity distributions of HLG modes in Fig. 1 (see also section III). From the expression obtained it follows that an LG beam with a positive topological charge is converted into an HG beam in such a way that the $n+l$ intensity zeros are located along the $\varphi=135°$ direction whereas the $n$ intensity zeros are ranged along the $\varphi=45°$ direction, while for a beam with a negative topological charge we have the $n$ intensity zeros along the $\varphi=135°$ direction, and the $n+l$ zeros along the $\varphi=45°$ direction. The astigmatic transformation of the LG beam by a cylindrical lens is relatively easy to implement in practice and can serve as an example of the express method for determining the TC.

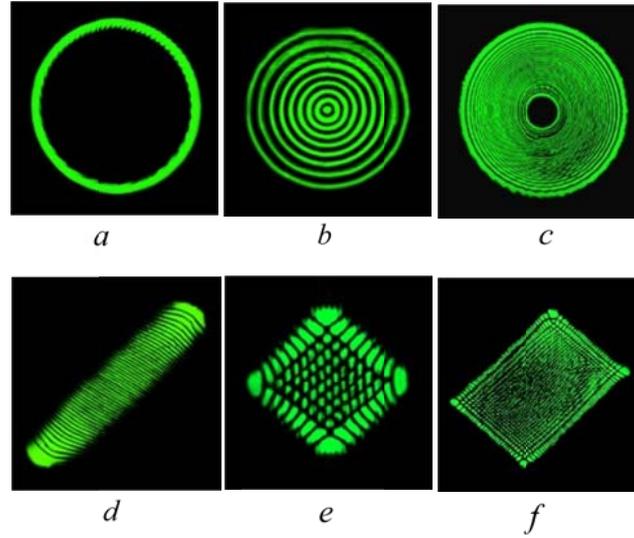

Fig. 2. Intensity distributions LG beams (a,b,c) at the initial plane $z=0$ and their astigmatic conversion into HG beams(d,e,f) at the $z=2f$ plane: (a) $l=-50, n=0$, (b) $l=0, n=8$, (c) $l=-25, n=25$, $f=12cm$

In our experiment, we used a cylindrical lens with a focal length of $f=12cm$ (see Section III). A simple calculation of the intensity zeros $n$ and $n+l$ in the intensity distribution of the HG beams in Fig. 2(b,d,e) along the $\varphi=45°$ and $\varphi=135°$ directions, respectively, allows one to find the TC $l$ and the number $n$ of the degenerate ring dislocations. The measurement of the TC and ring dislocations can also be performed using a spherical lens, if one turns it at a certain angle relative to the optical axis, as was done in Ref. [18] but at the same time, the Hermite polynomial argument remains a complex value since the parameter $c_y$ in Eq. (5) is not zero, and the intensity zeros are located strictly along the $\varphi=45°$ and $\varphi=135°$ directions. However, irregularities in the intensity distribution of the HG beams depicted in Fig. 2(d,f,e) indicate a complex mode composition, despite a certain topological charge, which determines its fine structure. It is this problem we will consider in the next section.

### III The fine mode structure of the LG beam and its measuring

In this section, we touch on the problem of measuring the fine structure of perturbed LG beams via the intensity moments technique in the basis of the HG modes. The approach of the intensity moments technique recommended itself well when describing perturbations of vortex beams by various opaque obstacles, where it was sufficiently to take into account only the birth-annihilation events of optical vortices with different TC, without addressing the modes with different initial phases and radial numbers. However, as we mentioned above, small perturbations, for example, of the holographic grating or inaccurate matching the laser beam and holographic element, can cause significant distortions of higher order vortex beams, although their TC does not change. A simple model of such a perturbation can serve a variation of amplitudes and initial phases of the HG modes in the LG beam spectrum.

Let us represent the complex amplitude of the LG beam as the sum of the HG modes [19]

$$\mathrm{LG}_{n,\pm l}(\mathbf{r}) = \frac{(-1)^n}{2^{2n+3l/2} n!} \times$$
$$\times \sum_{k=0}^{2n+l} (\pm 2i)^k P_k^{(n+l-k,n-k)}(0) \mathrm{HG}_{2n+l-k,k}(\mathbf{r}), \tag{10}$$

where

$$P_k^{(n-k,m-k)}(0) = \frac{(-1)^k}{2^k k!} \frac{d^k}{dt^k}\left\{(1-t)^n (1+t)^m\right\}\bigg|_{t=0} \tag{11}$$

is the Jacobi polynomial with a zero argument. The expression (10) describes the LG beam with the TC $\pm l$. A variation of the TC sign corresponds to a change in the sign before the imaginary unit on the right-hand side of (10). We also note that the LG and HG modes in Eq. (10) are not normalized. Also we will consider such perturbations of the LG beam that leave the composition of the HG modes in the superposition (10) constant. We chose two perturbation models that satisfy this requirement. The first model, a hard perturbation, allows one to sharply change the amplitudes of the HG modes and has the form

$$LG_{n,\pm l}(\mathbf{r},\varepsilon,\nu) = \frac{(-1)^n}{2^{2n+3l/2} n!} \times$$
$$\times \sum_{k=0}^{2n+\ell} \left(1 - \varepsilon P_k^{(n+l-k,n-k)}(\nu)\right) \times$$
$$\times (\pm 2i)^k P_k^{(n+\ell-k,n-k)}(0) \mathrm{HG}_{2n+\ell-k,k}(\mathbf{r}), \tag{12}$$

where $\varepsilon$ and $\nu$ are the control parameters. The second model describes a soft perturbation that weakly transforms the HG amplitudes when changing the perturbation:

$$LG_{n,\pm l}^{(\mathrm{pert})}(\mathbf{r},\varepsilon) = \frac{(-1)^n}{2^{2n+3l/2} n!} \sum_{k=0}^{2n+l} \varepsilon_k \times$$
$$\times (\pm 2i)^k P_k^{(n+l-k,n-k)}(0) HG_{2n+l-k,k}(\mathbf{r}), \tag{13}$$

where

$$\varepsilon_k = 1 - \varepsilon(-1)^k \sin\frac{\pi k}{2} \tag{14}$$

Typical intensity distributions in the hard mode perturbation are shown in Fig. 3. In this case, the second term in Eq. (12) suppresses quickly the first term, and the wave system responds weakly to perturbations.

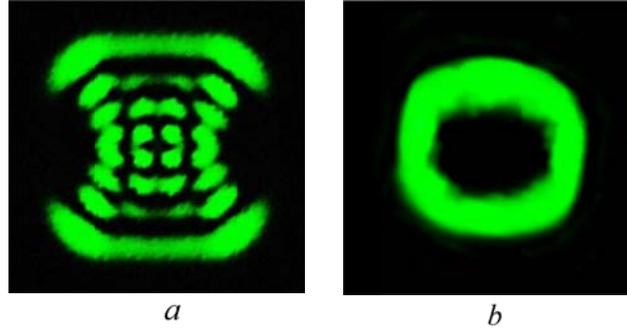

Fig. 3. Intensity distributions of the LG beam with $l=10, n=0$ for the hard perturbation with (a) $\varepsilon=0,1$, $v=0,2$ and (b) $\varepsilon=0,1$, $v=0,4$

The soft characteristics of the second model (13) turned out to be more acceptable for our consideration since the system responds to a perturbations in a wide range of changing the control parameter $\varepsilon$. The variety of perturbation options for this case depicts Fig. 4.

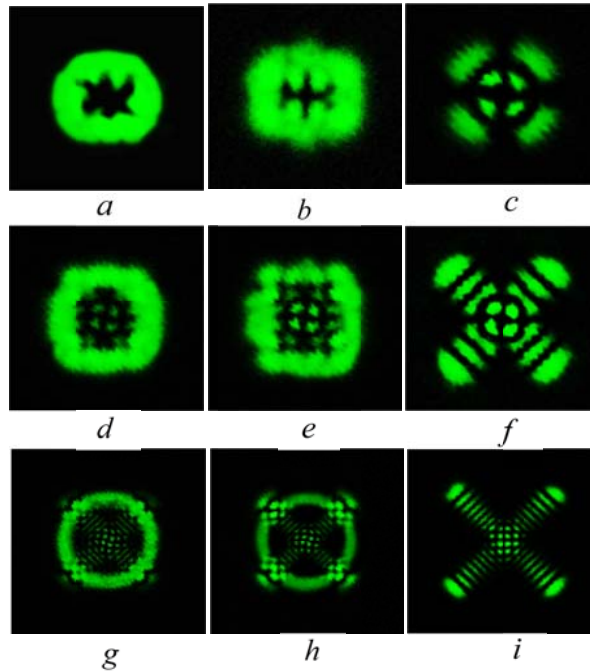

Fig. 4. Soft perturbation of LG beam for: (a,b,c) - $l=n-m=4$; (d,e,f) - $l=8$; (g,h,i) $l=20$: (a,d,g) - $\varepsilon_0=0.5$; (b,e,h) - $\varepsilon_0=1$ and (c, f, i) - $\varepsilon_0=50$, and $m=0$

It is convenient to distinguish three regions of perturbation: 1) small values of the parameter $\varepsilon \ll 1$ when the perturbation splits a higher-order axial vortex into single-charged vortices (Fig. 4(a,d,g)); 2) the option $\varepsilon \sim 1$ when topological dipoles (vortices with opposite signs of TC) are born and annihilated so that the OAM changes rapidly, and the TC of the entire beam can change sign (Fig. 4(b,d,h)); 3) the option $\varepsilon \gg 1$ when the main contribution is made by the second term in (14) and the OAM changes only slightly (Fig. 4 (c,f,i)). A visual representation of the perturbation options is illustrated in Fig. 5 by the example of diffraction gratings on the working cell of a spatial light modulator (SLM) that restores the LG beam with TC $l=8$.

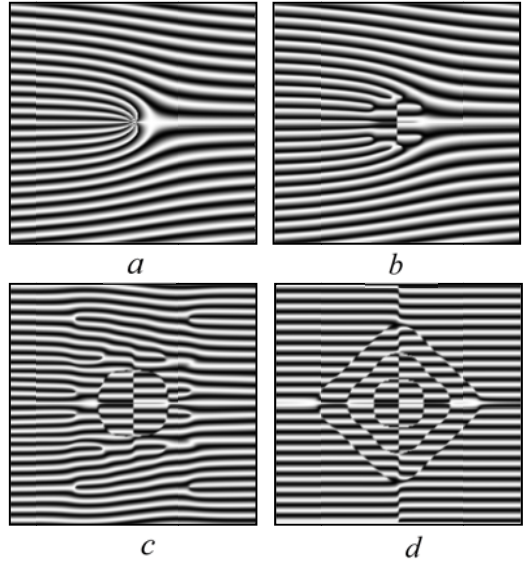

Fig. 5. Diffraction gratings of the perturbed LG beams: (a) $\varepsilon=0$, (b) $\varepsilon=0.01$, (c) $\varepsilon=1$, (d) $\varepsilon=20$

We note that the perturbation is located in the vicinity of the vortex core. The ratio of the disturbance area to the working area of the SLM matrix varied from 0.01 to 0.04. A small perturbation splits a high-order optical vortex into single vortices (Fig. 5(b)). Growth of the perturbation (Fig. 5(c)) leads to shaping topological dipoles, and for a large perturbation parameter (Fig. 5(d)), a system of figured edge dislocations are structured in the form of embedded squares and two edge dislocations along the x and y axes.

Now we turn to the theoretical interpretation of the problem of recording the HG mode spectra in the case of the soft perturbation options. As in the case of measuring the LG mode spectra [8,9] in the LG mode basis, in the presented case, we also analyze the LG mode intensity distribution but in the basis of the HG modes:

$$LG_{n,l}^{(ex)}(\mathbf{r}, z=0) = \sum_{k=0}^{2n+l} \varepsilon_k e^{i\beta_k} C_{2n+l-k,k} HG_{2n+l-k,k}(\mathbf{r}),$$

$$\Im_{n,\ell}(\mathbf{r}) = \left| LG_{n,\ell}^{(ex)}(\mathbf{r}, z=0) \right|^2, \qquad (15)$$

where $C_{2n+\ell-k,k}$ and $\beta_k$ are real mode amplitudes and initial phases, respectively.

For analysis, we present the intensity distribution (15) in terms of intensity moments in accordance with our work [21].

$$J_{p,q} = \frac{1}{J_{00}} \iint_{\mathbb{R}^2} M_{p,q} \Im_{n,\ell}(x,y) dx dy, \qquad (16)$$

where $M_{p,q}$ stands for a moment function, $J_{00} = \sum_{k=0}^{2n+\ell} |C_{2n+\ell-k,k}|^2 j_k$ is a total intensity of the LG beam and $j_k$ are elements of the HG basis. The intensity moments are measured experimentally so that $J_{p,q}$ are real values. The moment function should be chosen in such a way that from Eq. (16) it would be possible to obtain two independent systems of linear equations for the squared amplitudes and the cross terms for different numbers p and q. It can be shown that the function

$$M_{p,q} = H_p(\sqrt{2}x)H_q(\sqrt{2}y) \tag{17}$$

meets this requirement. In such a representation of the moment functions, the elements of the HG basis are specified by the integral

$$j_n = \int_{\mathbb{R}} e^{-2x^2} H_p(\sqrt{2}x) H_m(\sqrt{2}x) H_n(\sqrt{2}x) dx =$$

$$= \frac{2^{(p+m+n-1)/2} \sqrt{\pi} p! m! n!}{\left(\frac{1}{2}N - p\right)! \left(\frac{1}{2}N - m\right)! \left(\frac{1}{2}N - n\right)!}, \tag{18}$$

where $N = m + n + p = 0, 2, 4, \ldots$ and the indices m, n, p satisfies the triangle inequality (i.e., the sum of any two indices cannot be greater than the third). In other cases, the integral vanishes. For the beam intensity $J_{00}$ with $p = 0$, Eq. (18) turns into the orthogonality condition of the HG modes.

Let us consider in details the process of measuring the perturbed LG beams spectra with a topological charge $\ell = 4$ and zero radial number $n = 0$. As we have already noted, the intensity moments (16) are measured experimentally while the elements $j_k$ of the HG basis are easy calculated. Therefore, the problem of finding the spectral amplitudes $C_{2n+\ell-k,k}$ and their phases $\beta_k$ is reduced to writing a set of $2(\ell+1)$ linear equations for the amplitudes and cross terms with sinus and cosines. It should be noted that the equations for the squared amplitudes and phases can be separated. However, when solving equations for the phase difference, it is necessary to take into account the squared amplitudes. In order to write the equations for the squared amplitudes, we choose only even indices $p = 2, 4, 6, 8$, $q = 0$ and $p = 0$, $q = 2, 4, 6, 8$. As a result, we obtain a system of linear equations

$$J_{0,2} = 16C_{0,4}^2 + 12C_{1,3}^2 + 8C_{2,2}^2 + 4C_{3,1}^2,$$
$$J_{0,4} = 48C_{2,2}^2 + 144C_{1,3}^2 + 288C_{4,0}^2,$$
$$J_{0,6} = 960C_{1,3}^2 + 3840C_{0,4}^2,$$
$$J_{0,8} = 2680C_{0,4}^2,$$
$$J_{8,0} = 2680C_{4,0}^2,$$
$$J_{6,0} = 3840C_{4,0}^2 + 960C_{3,1}^2, \tag{19}$$

The obtained set of equations enables us to find all the squared amplitudes. To determine the cross terms with sinus and cosines of the phase differences $\beta_{k,j} = \beta_k - \beta_j$, we choose only even and only odd $p$ and $q$ indices

$$J_{2,2} = 48C_{1,3}^2 + 65C_{2,2}^2 + 48C_{3,1}^2 - 16\sqrt{6}C_{4,0}C_{2,2}\cos\beta_{0,2} +$$
$$+ 48C_{1,3}C_{3,1}\cos\beta_{1,3} + 16\sqrt{6}C_{0,4}C_{2,2}\cos\beta_{2,4}$$
$$J_{4,2} = 384C_{2,2}^2 + 576C_{3,1}^2 + 256\sqrt{6}C_{4,0}C_{2,2}\cos\beta_{0,2} +$$
$$+ 384C_{1,3}C_{3,1}\cos\beta_{1,3},$$
$$J_{6,2} = 3840C_{3,1}^2 + 1920\sqrt{6}C_{4,0}C_{2,2}\cos\beta_{0,2},$$
$$J_{7,1} = 13440C_{3,1}C_{4,0}\cos\beta_{0,1}.$$
(20)

The solution of the equations (20) has some specifics, since the equations contain both squared amplitudes and the amplitudes themselves. We chose the amplitudes of the HG modes as positive real quantities; therefore, they can be found from equations (19) while the phases $\beta_k$ take into account the signs of the amplitudes. In order to find phases from equations (20), it is enough to put $\beta_0 = 0$ from where we find the rest initial phases $\beta_1, \beta_2, \beta_3$ and $\beta_4$.

The experimental set-up was in details described in our articles [14] (Fig. 1) and [22] (Fig. 10). We employed the same set-up in our measurements. The analysis of the intensity moments was carried out in the beam waist plane. Figure 6 shows typical spectra of squared amplitudes and initial phases.

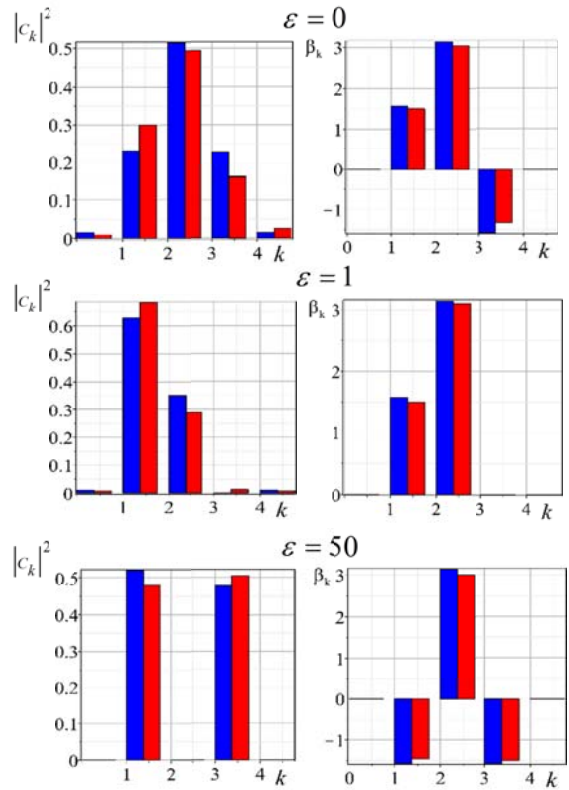

Fig. 6. Squared amplitudes $|c_k|^2$ and initial phases $\beta_k$ of HG mode in the LG node spectrum c $n = 0, m = 4$, blue -theory, red - experiment.

For comparison, the theoretical spectra calculated in accordance with the expression (13) are also plotted there. A slight mismatch between theory and experiment demonstrates the relative measurement error $\overline{\Delta C^2} \approx 3\%$, $\overline{\Delta \beta^2} \approx 6\%$. It is worth noting that the error in the measurements of the initial phases substantially depends on the accuracy of the measurements of the squared amplitudes spectrum. In fact, the presented measurements perform the digital sorting of the HG modes taking into account their amplitudes and initial phases. If necessary, one can restore the entire perturbed beam, getting rid of spurious noise.

## IV Analysis of the HG spectra

An important step in the analysis of the HG modes spectra is their matching with the usual spectra of the LG modes, the OAM, and estimation of the TC. In the Appendix we showed how to rewrite the HG mode spectrum (13) in terms of the LG modes:

$$LG_{n,\pm l}^{(\text{pert})}(\mathbf{r},\varepsilon) = LG_{n,\pm l}(\mathbf{r}) \mp \varepsilon \cdot \frac{(-1)^n e^{\pi i l/4}}{2^{2n+3l/2} n!} \times$$
$$\times \sum_{k=0}^{2n+l} (-2i)^k P_k^{(n+l-k,n-k)}(0) \times$$
$$\times (-1)^{\min} 2^{\max} (\min)! \cdot \text{Im}\{LG_{\min,2n+l-2k}(\mathbf{r})\}, \quad (21)$$

where $\min = \min(2n+l-k, k)$ and $\max = \max(2n+l-k, k)$.

For example, for $n=0$ we have

$$LG_{0,1}^{(\text{pert})}(\mathbf{r},\varepsilon) = e^{-r^2}(x + i[1-\varepsilon]y),$$
$$LG_{0,2}^{(\text{pert})}(\mathbf{r},\varepsilon) = e^{-r^2}(x^2 - y^2 + 2i[1-\varepsilon]xy),$$
$$LG_{0,3}^{(\text{pert})}(\mathbf{r},\varepsilon) = e^{-r^2}\big((x+iy)^3 -$$
$$- \tfrac{1}{2}i\varepsilon(3x^2 - y^2 + 3y[r^2 - 1])\big). \quad (22)$$

For the first two beams, a perturbation $\varepsilon = 1$ results in the switching of the TC and OAM signs. In the third beam, the signs of the TC and OAM are switched at the perturbation $\varepsilon = 2$, but for large values $\varepsilon$, the OAM does not reach its maximum value -3, as in the two previous cases, but the OAM tends to zero due to the competition of three terms in Eq. (22). The calculation of the OAM in the basis of the HG modes is easy to carry out using the standard approach [9,18] and Eq. (21) for the known LG mode amplitudes. Using (A5) and (A6) in the Appendix, we write the theoretical specific OAM for $n > 0, m \geq 2$ in the form

$$\ell_z = \frac{L_z}{J_{00}} = \frac{m + 2\varepsilon\left(\frac{\Lambda_{n-1} + \Lambda_{n+1}}{2^{n+m/2}}\right) \sin\frac{\pi(2n+m)}{4}}{1 - 2\varepsilon \frac{\lambda_n}{2^{n+m/2}} \sin\frac{\pi(2n+m)}{4} + \frac{\varepsilon^2}{2}} \quad (23)$$

On the other hand, the expression (15) enables us to tie together the OAM and the mode amplitudes. For example, the vortex beam $LG_{0,3}^{(\text{pert})}(\mathbf{r},\varepsilon)$ has the total intensity and OAM in the form

$$J_{00} = C_{1,2}^2 + 4C_{2,1}^2 + 3C_{3,0}^2, \quad (24)$$

$$L_z = C_{1,2}C_{2,1}\sin(\beta_1 - \beta_2) - 12C_{2,1}C_{3,0}\sin(\beta_1 - \beta_0). \quad (25)$$

The initial phases and amplitudes are found from the spectra of the HG modes, then from Eq. (24) and Eq. (25) we obtain the OAM per photon $\ell_z$. Typical theoretical OAM curves and the corresponding experimental points (circles, triangles, and squares) are shown in Fig. 7. As expected, the action of the perturbation on the LG beams is divided into two groups. The first group includes beams with a topological charge $|l| \leq 3$. The second group includes all other beams $|l| > 3$. From Fig. 7(a) it can be seen that the orbital momentum for $l_z = 0$, $l = 1, l = 2$ at $\varepsilon = 1$ and for $l = 3$ the OAM vanishes. At the same time, growing the perturbation leads to switching the signs of both the TC and OAM. The sign switching of the TC in these beams is easily estimated from Eq. (22). The first term in Eq. (21) makes the main contribution to the TC until the second term exceed the amplitude of the unperturbed LG beam. However, the sign of the second term controls the factor $e^{\pi i \ell/4}$ in such a way that the sign remains positive for $l > 4$, while the OAM will gradually decrease for any perturbations $\varepsilon$.

The OAM measurements were carried out in two ways. On the one hand, expression (15) allows us to calculate the field $LG_{n,l}^{(ex)}$ using the spectrum of amplitudes and phases of the HG mode. Then, using the expression

$$L_z = \operatorname{Im} \iint_{\mathbb{R}^2} LG_{n,l}^{(ex)*} \left( x \frac{\partial LG_{n,l}^{(ex)}}{\partial y} - y \frac{\partial LG_{n,l}^{(ex)}}{\partial x} \right) dxdy, \qquad (26)$$

we find the OAM. On the other hand, our experimental setup allows to perform independent measurements of the OAM using the intensity moments technique [14]. Therefore, in the case of deviation of the OAM measurement results in the first case by more than 5%, the measurements are duplicated by the direct measurement of the OAM on the basis of the intensity moments. As can be seen from Fig. 7, the deviation of experimental points from theoretical curves does not exceed 5% in the region $\varepsilon \in (0,5)$.

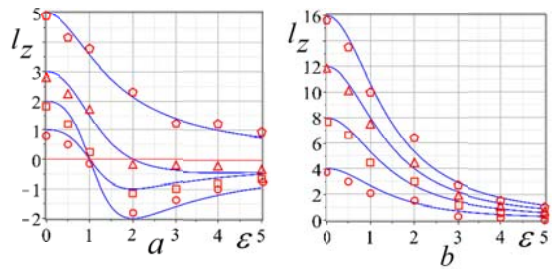

Fig. 7. Dependence of the OAM per photon $l_z = L_z / J_{00}$ on the control parameter $\varepsilon$ for perturbed LG beams with for (a) $l = 1$, $l = 2$, $l = 3$, $l = 5$; (b) $l = 4$, $l = 8$, $l = 12$, $l = 16$

## Conclusions

A new method for measuring the amplitudes and initial phases spectra of perturbed LG beams in the basis of HG modes was developed and implemented based on the approach of intensity moments. We have also theoretically substantiated and experimentally implemented a method for measuring the TC of Laguerre-Gaussian beams with a different number of ring dislocations. We have chosen such a form of the perturbation that distorts the high-order LG beam structure only in the vicinity of the optical vortex core generating a fine mode spectral structure. As the perturbation grows, the distortion region increases, but does not exceed 4% of the total area of the holographic grating on a spatial light modulator. Large perturbations promote appearing figured edge dislocations surrounding a high-order vortex core and two edge dislocations along the x and y axes. Such perturbations can be easily simulated by weak variations of the amplitudes and phases of the HG modes in the expansion of the LG beam field. We also theoretically substantiated and experimentally implemented a method for measuring the TC of Laguerre-Gaussian beams with a different number of ring dislocations. Using specific examples, an algorithm for measuring the spectrum of amplitudes and initial phases of the HG modes, OAM, and TC is demonstrated. General expressions are obtained for calculating the spectrum of HG modes, and with its help we found relations for estimating OAM. The presented theoretical and experimental studies make it possible to significantly expand the of intensity moments technique by measuring vortex beams both in the LG mode basis and in the HG mode basis in the case of various types of the beam perturbations.

Appendix

Let us simplify Eq. (13), substituting the representation of two terms (14) instead the perturbation $\varepsilon_k$. Obviously, the first (single) term contributes $LG_{n,\pm\ell}^{(\text{pert})}(\mathbf{r},\varepsilon)$ in the form of the initial unperturbed LG mode, since when $\varepsilon = 0$, the expression (13) turns into Eq. (10). Therefore

$$LG_{n,\pm l}^{(\text{pert})}(\mathbf{r},\varepsilon) - LG_{n,\pm l}(\mathbf{r}) = \frac{(-1)^n}{2^{2n+3l/2}n!} \times$$

$$\times \sum_{\substack{k=0 \\ k \text{ odd}}}^{2n+l} (-1)^{(k+1)/2} \varepsilon \cdot (\pm 2i)^k P_k^{(n+l-k,n-k)}(0) HG_{2n+l-k,k}(\mathbf{r}) =$$

$$= \varepsilon \cdot \frac{(-1)^n}{2^{2n+3l/2}n!} \times$$

$$\times \sum_{0 \leq 2k+1 \leq 2n+l} (-1)^{k+1} \cdot (\pm 2i)^{2k+1} P_{2k+1}^{(n+l-2k-1,n-2k-1)}(0) HG_{2n+l-2k-1,2k+1}(\mathbf{r}) =$$

$$= \mp i\varepsilon \cdot \frac{(-1)^n}{2^{2n+3l/2}n!} \sum_{0 \leq 2k+1 \leq 2n+l} 2^{2k+1} P_{2k+1}^{(n+l-2k-1,n-2k-1)}(0) HG_{2n+l-2k-1,2k+1}(\mathbf{r}),$$  (A1)

where we used the oddness condition of the summation index k, making the substitution $k \to 2k+1$. We see now that the perturbing component is always imaginary. Next, we make the reverse substitution, and get rid of the oddness condition of the summation index k by adding an additional factor:

$$LG_{n,\pm l}^{(\text{pert})}(\mathbf{r},\varepsilon) - LG_{n,\pm l}(\mathbf{r}) = \mp i\varepsilon \cdot \frac{(-1)^n}{2^{2n+3l/2}n!} \times$$

$$\times \sum_{0 \leq k \leq 2n+l} \frac{1-(-1)^k}{2} \cdot 2^k P_k^{(n+l-k,n-k)}(0) HG_{2n+l-k,k}(\mathbf{r}) =$$

$$= \mp i\varepsilon \cdot \frac{(-1)^n}{2^{2n+3l/2+1}n!} \sum_{k=0}^{2n+\ell} \left\{ 2^k - (-2)^k \right\} P_k^{(n+l-k,n-k)}(0) HG_{2n+l-k,k}(\mathbf{r}) =$$

$$= \mp i\varepsilon \cdot \frac{(-1)^n}{2^{n+1}n!} \cdot \left\{ HG_{n+l,n}\left(\frac{x+y}{\sqrt{2}},\frac{x-y}{\sqrt{2}}\right) - HG_{n+l,n}\left(\frac{x-y}{\sqrt{2}},\frac{x+y}{\sqrt{2}}\right) \right\},$$  (A2)

where was employed the expression (3.10) from Ref. [23]:

$$\sum_{k=0}^{n+m} (\pm 2)^k P_k^{(n-k,m-k)}(0) HG_{n+m-k}(x) HG_k(y) =$$
$$= 2^{(n+m)/2} HG_n\left(\frac{x \pm y}{\sqrt{2}}\right) HG_m\left(\frac{x \mp y}{\sqrt{2}}\right).$$

To find the difference in the braces of (A2), we use the expression (4.20) from Ref. [23]

$$HG_{n+\ell,n}\left(\frac{x-y}{\sqrt{2}}, \frac{x+y}{\sqrt{2}}\right) = \frac{e^{\pi i \ell/4}}{2^{n+\ell/2}} \times$$
$$\times \sum_{k=0}^{2n+\ell} (-2i)^k P_k^{(n+\ell-k, n-k)}(0) \cdot (-1)^{\min} 2^{\max} (\min)! \cdot LG_{\min, 2n+\ell-2k}(x,y). \quad (A3)$$

Then we finally obtain

$$LG_{\min, 2n+l-2k}(x,y) - LG^*_{\min, 2n+l-2k}(x,y) =$$
$$= 2i \cdot \operatorname{Im} LG_{\min, 2n+l-2k}(x,y), \quad (A4)$$

that leads to Eq.(21). Using (A5) and omitting the intermediate calculations, we write the expressions for the intensity $J_{00}$ and OAM $L_z$ for $n > 0, m \geq 2$ in the form

$$J_{00}(\varepsilon) = \frac{\pi}{2} \frac{(n+m)!}{2^m n!} \left\{1 - 2\frac{\lambda_n}{2^{n+m/2}} \sin\frac{\pi(2n+m)}{4} + \frac{\varepsilon}{2}\right\}, \quad (A5)$$

$$L_z(\varepsilon) = \frac{\pi}{2} \frac{(n+m)!}{2^m n!} \times$$
$$\times \left\{m + 2\varepsilon \left(\frac{\Lambda_{n-1} + \Lambda_{n+1}}{2^{n+m/2}}\right) \sin\frac{\pi(2n+m)}{4}\right\}, \quad (A6)$$

where $\Lambda_{n-1} = (n+m+1)\lambda_{n-1}$, $\Lambda_{n+1} = (n+1)\lambda_{n+1}$, $\lambda_k = (-2)^k P_k^{(n+m-k,n-k)}(0)$.

**Funding**

The reported study was funded by RFBR according to the research project № 19-29-01233.

**Disclosures**

The authors declare no conflicts of interest.